\documentclass[aps,prl,reprint,showpacs,floatfix]{revtex4-1}

\usepackage{color}

\usepackage[utf8]{inputenc}
\usepackage{amsmath,amssymb,amstext}
\usepackage{amsthm}
\usepackage{dsfont}
\usepackage{graphicx}

\def\sout{\bgroup\markoverwith
{\textcolor{red}{\rule[0.5ex]{2pt}{0.5pt}}}\ULon}
\def\be{\begin{equation}}
\def\ee{\end{equation}}
\def\bes{\begin{equation*}}
\def\ees{\end{equation*}}
\def\bea{\begin{eqnarray}}
\def\eea{\end{eqnarray}}
\def\beas{\begin{eqnarray*}}
\def\eeas{\end{eqnarray*}}
\def\bal#1\eal{\begin{align}#1\end{align}}
\def\bals#1\eals{\begin{align*}#1\end{align*}}
\newcommand{\bra}[1]{\langle #1|}
\newcommand{\ket}[1]{|#1\rangle}
\newcommand{\braket}[2]{\langle #1|#2\rangle}

\usepackage[normalem]{ulem}

\usepackage{braket}

\graphicspath{{figures/}}

\makeatletter
\AtBeginDocument{\let\LS@rot\@undefined}
\makeatother

\begin{document}

\title{Formally Self-Adjoint Hamiltonian for the Hilbert–Pólya Conjecture}

\author{Enderalp Yakaboylu}
\email{enderalp.yakaboylu@mpq.mpg.de}
\affiliation{Max Planck Institute of Quantum Optics, 85748 Garching, Germany}

\date{\today}

\begin{abstract}

We construct a formally self-adjoint Hamiltonian whose eigenvalues correspond to the nontrivial zeros of the Riemann zeta function. We consider a two-dimensional Hamiltonian which couples the Berry-Keating Hamiltonian to the number operator on the half-line via a unitary transformation. We demonstrate that the unitary operator, which is composed of squeeze (dilation) operators and an exponential of the number operator, confines the eigenfunction of the Hamiltonian to one dimension as the squeezing parameter tends towards infinity. The Riemann zeta function appears at the boundary of the resulting confined wave function and vanishes as a result of the imposed boundary condition. If the formal argument presented here can be made more rigorous, particularly if it can be shown rigorously that the Hamiltonian remains self-adjoint under the imposed boundary condition, then our approach has the potential to imply that the Riemann hypothesis is true.

\end{abstract}

\maketitle

The Riemann hypothesis, which requires no introduction, is considered to be one of the most important unsolved problems in mathematics, if not the most. The hypothesis states that every nontrivial zero of the Riemann zeta function $\zeta (s)$ lies on the critical line $\text{Re}(s) = 1/2$. The Riemann zeta function, which can be written as
\bes
\zeta(s) = \sum_{m=0}^\infty \frac{1}{(m+1)^{s}} \, ,
\ees
for $\text{Re}(s) > 1$, can be defined for other complex values of $s$ via analytic continuation. Particularly, for $\text{Re}(s) > 0$, (which we consider hereafter) it can be  given via the Dirichlet eta function
\bals
\nonumber \left(1 - 2^{1-s} \right) \zeta(s) & = \eta(s) = \sum_{m=0}^\infty \frac{(-1)^m}{(m+1)^s} \\
\label{zeta}
 & = \frac{1}{\Gamma(s)} \int_0^\infty d u \, u^{s-1} \frac{e^{-u}}{1+ e^{-u}} \, .
\eals

Despite strong numerical evidence of its validity-- it is known to be true for the first $10^{13}$ zeros~\cite{gourdon20041013}, and various analytical attempts, the hypothesis has not yet been proven. Among several pure mathematical approaches to establishing the Riemann hypothesis, there is one particular approach that has also attracted physicists; Hilbert–Pólya conjecture. It asserts that the imaginary parts of the nontrivial zeros of the Riemann zeta function correspond to eigenvalues of a self-adjoint Hamiltonian. Since the works of Selberg~\cite{selberg1956harmonic} (the Selberg trace formula), and Montgomery~\cite{montgomery1973pair} and Odlyzko~\cite{odlyzko1987distribution} (the connection between the distribution of the nontrivial zeros and the eigenvalues of a random Hermitian matrix drawn from the Gaussian unitary ensemble), the Hilbert–Pólya conjecture has a more solid basis. The research on the field has further intensified with the Berry-Keating program. 

Berry and Keating conjectured that the classical counterpart of such a Hamiltonian would have the form $x p$~\cite{berry1999h,berry1999riemann}. This simple suggestion was based on rather a heuristic and semiclassical analysis. In fact, such a classical Hamiltonian was previously introduced by Connes in Ref.~\cite{connes1999trace}, where the nontrivial zeros appear as missing spectral lines in a continuum. As a quantum counter part, Berry and Keating considered the simplest  formally self-adjoint Hamiltonian, $\hat{H}^\text{BK} = \left( \hat{x} \, \hat{p}  + \hat{p} \,  \hat{x} \right)/2 $, which is also referred to as the generator of dilation, $\hat{H}^\text{BK} \equiv \hat{D}$. Although the Berry-Keating Hamiltonian corresponds to the correct structure of the eigenvalues conjectured by Hilbert and Pólya, and provides the average number of nontrivial zeros up to a given height, this Hamiltonian is still quite far from being concrete. More explicitly, it is not known how to reveal the Riemann zeta function in the corresponding eigenvalue equation, and it is not clear on which space this Hamiltonian should act in order to get the  condition imposing the Riemann zeta function to be zero, for more details see the works of Sierra~\cite{sierra2007h,sierra2019riemann}.

An important advance in the Berry-Keating program has recently been put forward by Bender, Brody, and M\"{u}ller~\cite{bender2017hamiltonian}. They introduced a Hamiltonian that is a similarity transformation of the Berry-Keating Hamiltonian. The eigenvalue solutions are given in terms of the Hurwitz zeta function with a boundary condition, yielding the Riemann zeta function. However, the Hamiltonian is not Hermitian and the imposed boundary condition is not consistent with the similarity transformation, which changes the domain of the wave function~\cite{bellissard2017comment,twamley2006quantum}.

In this Letter, we consider the following formally self-adjoint Hamiltonian
\be
\label{initial_ham}
\hat{H} = \hat{U} \left( \hat{H}^\text{BK}_x \otimes I_y + I_x \otimes \hat{N}_y \right) \hat{U}^\dagger \, ,
\ee
which is a unitary transformation of two self-adjoint operators; the Berry-Keating Hamiltonian, $\hat{H}^\text{BK}_x $, which acts on the Hilbert space $\mathcal{H}_x  = L^2[0,\infty) $ associated with the variable $x$, and the number operator, $\hat{N}_y$, defined on the Hilbert space $\mathcal{H}_y  =  L^2[0,\infty)$ with the variable $y$. $I$ is the identity operator in the corresponding space. We define the unitary operator as
\be
\label{unitary_operator}
\scalebox{0.93}{$
\hat{U} = e^{i \lambda \hat{D}_y } e^{-i \hat{D}_x \otimes \ln(\hat{N}_y + 1)} e^{i \pi \hat{N}_x/2} e^{i \hat{D}_x \otimes \ln(\hat{N}_y + 1) } e^{-i \lambda  \hat{D}_y }$} \, .
\ee
We will show that in the limit of the called squeezing parameter $\lambda$ going to infinity, the wave function is confined to the $x$-direction and the Riemann zeta function emerges at the boundary of the confined wave function. A natural boundary condition forces it to be zero, and hence the eigenvalues of the Hamiltonian~\eqref{initial_ham} as $\lambda \to \infty$ are given in terms of the nontrivial zeros of the Riemann zeta function.

In the proposed Hamiltonian~\eqref{initial_ham}, the Berry-Keating Hamiltonian, $\hat{H}^\text{BK}$, is well-defined and self-adjoint on the Hilbert space $L^2[0,\infty)$, which can be shown by comparing the dimensions of the called ``deficiency subspaces'' (both the deficiency indices are equal to zero)~\cite{bonneau2001self,faris2006self,twamley2006quantum}. 
Furthermore, it is the generator of the dilation operator $\hat{S} = e^{- i \lambda \hat{D}} $, recall that $\hat{H}^\text{BK} = \hat{D}$. This operator appears as a conformal transformation in conformal field theories and the unitary squeeze operator in quantum optics, where the parameter $\lambda$ labels the amount of squeezing. The squeeze operator  fulfills the properties, $\hat{S} \, \hat{p} \, \hat{S}^\dagger  = \hat{p} \, e^\lambda$, $\hat{S} \, \hat{x} \, \hat{S}^\dagger = \hat{x} \, e^{-\lambda} $, and
\be
\label{squeeze_rel}
\bra{x} \hat{S} \ket{\psi}  = e^{-\lambda/2} \psi (e^{-\lambda} x) \, .
\ee
The corresponding eigenvalue equation for $\hat{H}^\text{BK}_x = \left( \hat{x} \, \hat{p}_x  + \hat{p}_x \,  \hat{x} \right)/2 $, is given by $ \hat{H}^\text{BK}_x \ket{\phi_s} = i(s-1/2) \ket{\phi_s} $ with the generalized eigenfunction $  \phi_s (x) \equiv \braket{x| \phi_s}  = x^{-s}/\sqrt{2 \pi} $. As the eigenvalues of a self-adjoint Hamiltonian are real, each $s$ has to be a complex number with $\text{Re}(s) = 1/2.$  Although the form of the eigenvalues indicates a clue, the core part of the problem is how to identify these eigenvalues with the nontrivial zeros of the Riemann zeta function, which is the essential goal of the present Letter. 

The number operator in the Hamiltonian~\eqref{initial_ham} is defined on the half-line $y \in [0,\infty)$, as well, and reads
\bes
\hat{N}_y = \frac{1}{2}\left(\hat{y} \, \hat{p}_y^2 + \hat{p}_y^2 \, \hat{y} + \frac{\hat{y}}{2} - 1 \right) \, ,
\ees
with the eigenvalue equation $\hat{N}_y \ket{n_y} = n_y \ket{n_y}$ and $n_y = 0 , 1 \cdots \infty$. The eigenvalue equation is nothing but the Laguerre ODE, and hence the self-adjointness can be shown in the context of Sturm–Liouville theory. 
The corresponding eigenfunctions can be given in terms of the Laguerre polynomials; $  \chi_{n_y} (y) \equiv \braket{y|n_y}= e^{-y/2} L_{n_y} (y)$. 

Accordingly, the eigenstate of the Hamiltonian~\eqref{initial_ham} can be written as
\be
\label{eigenstate_0}
\ket{\Psi_{s,n_y}} = \hat{U} \ket{\phi_s} \otimes \ket{n_y}  \, ,
\ee
and the associated eigenvalues read $E_{s,n_y} = i(s-1/2) + n_y  $. The unitary operator defined in Eq.~\eqref{unitary_operator} is composed of unitary squeeze operators and a unitary  exponential of the number operator. While the first squeeze operator, $S_1 =  e^{-i \lambda  \hat{D}_y } $, squeezes the eigenstate~\eqref{eigenstate_0} in the $y$-direction with the amount of $\lambda$, the second one, $S_2 = e^{i \hat{D}_x \otimes \ln(\hat{N}_y + 1) }$, squeezes it in the $x$-direction with the amount depending on the energy level in the $y$-direction. Subsequently, the state accumulates a phase factor in the basis of the eigenstate of the number operator with $\hat{R} = e^{i \pi \hat{N}_x/2}$. Lastly, the eigenstate~\eqref{eigenstate_0} is dilated back with the action of the two inverse squeeze operators. For a shorthand notation, we define the unitary operator as $ \hat{U} = \hat{S}_1^\dagger \hat{S}_2^\dagger  \hat{R} \hat{S}_2 \hat{S}_1 $, and perform the action of each operator respectively, which will bring the construction of the Hamiltonian~\eqref{initial_ham} to light.

After we act the first squeeze operator, we expand the state $\hat{S}_1 \ket{n_y}$ in terms of the eigenstate of the number operator by using the identity operator $\sum_{m_y=0}^\infty \ket{m_y}\bra{m_y}$ in the corresponding space. Hence, the eigenstate~\eqref{eigenstate_0} is
\bes
\ket{\Psi_{s,n_y}} = \hat{S}_1^\dagger \hat{S}_2^\dagger \hat{R} \hat{S}_2 \sum_{m_y =0}^\infty \bra{m_y}\hat{S}_1 \ket{n_y} \ket{\phi_s} \otimes \ket{m_y}\, .
\ees
The second squeeze operator, which acts on both the Hilbert spaces $\mathcal{H}_x$ and $\mathcal{H}_y$, couples the Berry-Keating Hamiltonian to the transformed number operator, $\hat{S}_1 \hat{N}_y \hat{S}_1^\dagger$. The generators $\hat{D}_x$ and $\ln(\hat{N}_y + 1)$ pick the eigenvalues from the associated eigenstates so that
\bals
\ket{\Psi_{s,n_y}} & = \hat{S}_1^\dagger \hat{S}_2^\dagger \hat{R} \sum_{m_y =0}^\infty \bra{m_y}\hat{S}_1 \ket{n_y} (m_y + 1)^{-s+\frac{1}{2}} \\
& \times \ket{\phi_s} \otimes \ket{m_y}  \, .
\eals
Before we apply the exponential of the number operator $\hat{R}$, we expand the state $\ket{\phi_s}$ in terms of the eigenstate of the number operator $\hat{N}_x$ in $\mathcal{H}_x$. Then, it follows that
\bals
\ket{\Psi_{s,n_y}} & = \hat{S}_1^\dagger \hat{S}_2^\dagger \sum_{m_y =0}^\infty \bra{m_y}\hat{S}_1 \ket{n_y}  (m_y + 1)^{-s+\frac{1}{2}} \\
\nonumber & \times   \sum_{m_x = 0}^\infty \braket{m_x | \phi_s} i^{m_x} \ket{m_x} \otimes \ket{m_y} \, .
\eals
Next, we act the remaining inverse squeeze operators
\bals
& \ket{\Psi_{s,n_y}} = \sum_{m_y=0}^\infty \bra{m_y}\hat{S}_1 \ket{n_y} (m_y + 1)^{-s+\frac{1}{2}}  \\
\nonumber & \times   \sum_{m_x =0}^\infty i^{m_x} \braket{m_x|\phi_s}  e^{-i \hat{D}_x \ln(m_y +1)} \ket{m_x} \otimes \hat{S}_1^\dagger \ket{m_y} \, .
\eals
Now, by using the property \eqref{squeeze_rel} and the relation
\bes
\bra{m_y}\hat{S}_1 \ket{n_y} =  e^{-\lambda/2} \int_0^\infty d y' \,  \chi_{m_y}^*(y') \chi_{n_y} (e^{-\lambda} y')\, ,
\ees
we define the position space wave function as
\bal
\nonumber \Psi_{s,n_y} & (x,y)  \equiv \bra{x}\otimes \braket{y|\Psi_{s,n_y}} = \int_0^\infty d y' \chi_{n} (e^{-\lambda} y')  \\  
\label{wave_function} & \times  \sum_{m_y =0}^\infty \frac{\chi_{m_y}^*(y') \chi_{m_y} \left(e^\lambda y \right)}{(m_y + 1)^{s}} \varphi_{s} \left(\frac{x}{m_y + 1} \right)   \, ,
\eal
where $ \varphi_{s} (x ) \equiv \bra{x}\hat{R} \ket{\phi_s}= \sum_{m_x =0 }^\infty \chi_{m_x} (x) \bra{m_x}\hat{R}\ket{\phi_s}$.

Here, we would like to comment on the resulting wave function~\eqref{wave_function}. As a matter of fact, the whole construction of the Hamiltonian~\eqref{initial_ham} stems from establishing the Riemann zeta term, $\sum_{m_y = 0}^\infty (-1)^{m_y}(m_y + 1)^{-s}$, inside the wave function. The first squeeze transformation allows us to consider the superposition of the eigenstate of the number operator, which results the sum. The second squeeze transformation, $\hat{S}_2$, generates the factor $(m_y + 1)^{-s+\frac{1}{2}}$. In order to eliminate the undesirable residual term, $\sqrt{m_y +1}$, we first change the basis with the operator $\hat{R}$, and then apply the inverse squeeze operator $\hat{S}_2^\dagger$. The undesirable term vanishes at the wave function level, which is implied by Eq.~\eqref{squeeze_rel}. The final inverse squeeze operator, which acts as a counter term to the exponential growth of the wave function resulting from the action of $\hat{S}_1$, is introduced in order to deduce the factor $(-1)^{m_y}$ in the limit of the squeezing parameter going to infinity, i.e., the integral in Eq.\eqref{wave_function} reads $\lim_{\lambda \to \infty}\int_0^\infty d y' \chi_{n_y} (e^{-\lambda} y') \chi_{m_y}^* (y') = 2 (-1)^{m_y}$. We remark that, instead of the exponential of the number operator $\hat{R}$, one could, very well, choose any other operator that does not commute with the dilation operator in $\mathcal{H}_x$. We would like to further point out that the factor, $(m_y + 1)^{-s}$, also would be obtained directly by applying the translation operator, $\hat{T} = \exp(i \hat{p}_x \otimes (\hat{N}_y + 1) ) $. Explicitly, we would have $\bra{x} \otimes \bra{y} \hat{T} \ket{\phi_s} \otimes \ket{m_y} = (x + m_y + 1)^{-s} \chi_{m_y} (y) /\sqrt{2 \pi}$ and would impose a boundary condition at $x=0$. However, the operator $\hat{T}$ alters the domain $x \in [0,\infty)$, which is a consequence of the fact that $\hat{p}_x$ is not a self-adjoint operator on the space $L^2[0, \infty)$~\cite{bonneau2001self,faris2006self}. In contrast to the translation operator, the squeeze operator $\hat{S}$ preserves the domain of the wave function, thus it is well-defined. We underline that such an explicit derivation of the wave function vis-\'a-vis the Riemann zeta function takes its source from the number operator $\hat{N}_y$, which couples to the Berry-Keating Hamiltonian via the unitary transformation. 

Now, we impose the Dirichlet boundary condition $\Psi_{s,n_y} (x =0,y) = 0$. Namely, we regard the  wave function $\Psi_{s,n_y} (x,y) $, which satisfies the corresponding differential equation for the Hamiltonian~\eqref{initial_ham} as well as the Dirichlet boundary condition, as the eigenfunction associated to the eigenvalues $E_{s,n_y}$. The motivation of such a boundary condition simply follows from the Hilbert space $L^2[0,\infty)$. Nevertheless, we should note that the imposed boundary condition might restrict the domain such that the Hamiltonian might not be self-adjoint anymore. 

For the sake of transparency, we first write the total wave function~\eqref{wave_function} as
\bals
 & \Psi_{s,n_y} (x,y)  = \frac{1}{\Gamma[s]} \int_0^\infty du \, u^{s-1} e^{-u} \int_0^\infty d y' \,  \chi_{n_y} (e^{-\lambda} y') \\
 & \times  \sum_{m_y = 0}^\infty \chi_{m_y}^*(y') \chi_{m_y} (e^\lambda y)  e^{-m_y u}  \, \varphi_{s} \left(\frac{x}{m_y + 1} \right) \, ,
\eals
where we used the identity
\bes
(m_y+1)^{-s} = \frac{1}{\Gamma(s)} \int_0^\infty du \, u^{s-1} e^{-(m_y+1)u} \, .
\vspace{0.3cm}
\ees
Then, by using the Mehler kernel formula for Laguerre polynomials~\cite{szeg1939orthogonal},
\bes
 \sum_{m=0}^\infty \chi_m^*(y') \chi_m (y) t^m  = \frac{\exp\left[-\frac{(y+y')}{2}\frac{1+t}{1-t}\right]}{1-t} I_0 \left( \frac{2 \sqrt{y y' t}}{1-t}\right) \, ,
\ees
with $I_n$ being the modified Bessel function of the first kind, we identify the wave function at the boundary as
\bal
\label{wf_at_bc}
\nonumber & \Psi_{s,n_y} (0,y) = \frac{\varphi_s (0)}{\Gamma[s]} \int_0^\infty du  \frac{u^{s-1}  e^{-u}}{1-e^{-u}} \int_0^\infty d y' \chi_{n_y} (e^{-\lambda} y') \\
 & \times \exp\left[-\frac{(e^\lambda  y + y')}{2}\frac{1+e^{-u}}{1-e^{-u}}\right] I_0 \left( \frac{2 \sqrt{e^\lambda y y' e^{-u}}}{1-e^{-u}}\right) \, ,
\eal
where $\varphi_s (0) = \Gamma[1-s](-2 i)^{\frac{1}{2}-s}/\sqrt{2 \pi}$. As we noted before, one can specify a different operator for $\hat{R}$ as long as the function $\varphi_s (0)$ does not vanish. 

We observe that in the limit of $\lambda \to \infty$, the total wave function at the boundary can be written simply as
\bals
\Psi_{s,n_y} (0,y)& = \frac{2 \varphi_s (0)}{\Gamma[s]} \int_0^\infty du \, u^{s-1} \frac{e^{-u}}{1+ e^{-u}} \\
\nonumber & \times \lim_{\lambda \to \infty}\exp\left(- \frac{(1-e^{-u}) y e^\lambda}{2(1+e^{-u})} \right) \, ,
\eals
which vanishes apart from the point $y=0$. Therefore, in the infinite limit, the wave function at the boundary is given in terms the Riemann zeta function
\be
\label{dirichlet_bc}
\Psi_{s,n_y} (0,y) = 2 \varphi_s (0) \delta_{y,0} (1-2^{1-s}) \zeta(s)\, ,
\ee
with $\delta_{y,0}$ being the Kronecker delta. Consequently, the Dirichlet boundary condition, which specifies the eigenvalues, can only be satisfied with $\zeta(s)  = 0$ for an arbitrary value of $y$. As the proposed Hamiltonian is formally self-adjoint, the eigenvalues are expected to be real, and hence the Riemann zeta function vanishes for the values of $\text{Re}(s) =1/2$.

We can render our result in a more intuitive and solid basis with a different perspective. We notice that in the limit of $\lambda \to \infty$, the wave function~\eqref{wave_function} vanishes, except at the point $y=0$. In other words, the unitary operator~\eqref{unitary_operator}, particularly the action of the squeeze operators $\hat{S}_1$ and $\hat{S}_1^\dagger$, confines the two-dimensional wave function to the $x$-direction as the squeezing parameter goes to infinity. The resulting confined one-dimensional wave function can be identified via $\Phi_s(x) \equiv  \Psi_{s,n_y} (x,0)$ as
\be
\label{1d_wave_func}
 \Phi_{s} (x) = 2 \sum_{m = 0}^\infty \frac{(-1)^m}{(m + 1)^{s}} \varphi_{s} \left(\frac{x}{m + 1} \right)   \, .
\ee
Then, by imposing the obvious boundary condition, $\Phi_s (0) = 0$, we show that the Riemann zeta function vanishes. However, we are not able to demonstrate rigorously either that the confined wave function~\eqref{1d_wave_func} is the eigenfunction of the corresponding one-dimensional confined Hamiltonian, resulting from $\lim_{\lambda \to \infty} H$, or that the confined Hamiltonian remains self-adjoint on the domain where the wave function~\eqref{1d_wave_func} is defined. Therefore, a more rigorous analysis is essential.

Nevertheless, for finite but sufficiently large values of $\lambda$, we can show that the Riemann hypothesis holds approximately true. Accordingly, instead of the unitary operator given in Eq.~\eqref{unitary_operator}, we consider its slightly modified version
\be
\label{unitary_operator_2}
\hat{\tilde{U}}  = e^{-2 i \lambda \hat{D}_y }e^{2 i \lambda \hat{D}_x } \hat{U}  \, .
\ee
By following the same steps we took as above, we define the wave function as
\bals
 \tilde{\Psi}_{s,n_y} (x,y)  &  = \int_0^\infty d y' \chi_{n} (e^{-\lambda} y')  \\  
\nonumber & \times  \sum_{m_y =0}^\infty \frac{\chi_{m_y}^*(y') \chi_{m_y} \left(e^{-\lambda} y \right)}{(m_y + 1)^{s}} \varphi_{s} \left(\frac{x e^{2 \lambda}}{m_y + 1} \right)   \, .
\eals
Then, the wave function at the boundary is given by
\bals
& \tilde{\Psi}_{s,n_y} (0,y) = \frac{2 \varphi_s (0)}{\Gamma[s]} \int_0^\infty du \, u^{s-1} \frac{e^{-u}}{1+ e^{-u}} \\
\nonumber & \times \left(1 - e^{-\lambda}(2n_y + 1 + y) \frac{1-e^{-u}}{1+e^{-u}} + O(e^{-2 \lambda})\right) \, ,
\eals
which can be further written in terms of the Riemann zeta function as
\bals
& \tilde{\Psi}_{s,n_y} (0,y) = 2 \varphi_s (0)  (1-2^{1-s}) \\
\nonumber & \times \left( \zeta(s)  + \frac{1}{e^{\lambda}}(2n_y + 1 + y) \xi(s)  + O(e^{-2 \lambda}) \right) \, ,
\eals
with $\xi(s) =\zeta(s) -2 (1-2^{2-s})\zeta(s-1)/(1-2^{1-s})$. Similar to the case of the unitary operator~\eqref{unitary_operator}, in the infinite limit of the squeezing parameter, the wave function at the boundary vanishes under the condition $\zeta(s) = 0$. For finite but sufficiently large values of $\lambda$, since the function $\xi(s)$ is finite for the nontrivial zeros, the wave function approximately vanishes  with the same condition.

In brief, we introduced a formally self-adjoint Hamiltonian, which is unitarily equivalent to the Berry-Keating Hamiltonian plus the number operator defined on the space $L^2[0,\infty)$. The unitary transformation, which couples the Berry-Keating Hamiltonian and the number operator, first squeezes the wave function, then rotates it, and finally dilates it back. In the limit as the squeezing parameter goes to infinity, the wave function is confined to one dimension and given in terms of the Riemann zeta function at the boundary so that the eigenvalues correspond to the nontrivial zeros under the imposed boundary condition. However, we need to underline that the imposed Dirichlet boundary condition might restrict the domain of the Hamiltonian~\eqref{initial_ham} such that its self-adjointness breaks down. Therefore, a more rigorous analysis should be undertaken for a concrete conclusion. 

As we explicitly demonstrated in two difference cases, one can further diversify the unitary transformation to establish the Berry-Keating Hamiltonian to the Riemann hypothesis by the virtue of the number operator defined on the half-line. For instance, we can modify the unitary operator in such a way that it confines the wave function to the $y$-direction, instead of to the $x$-direction, and then we impose a boundary condition for the resulting confined wave function in the $y$-direction. Furthermore, the factor $(-1)^{m_y}$ can also be obtained by the action of an additional exponential of the number operator in $\mathcal{H}_y$, which allows us to consider different variates of the unitary operator. Finally, we comment that through the instrumentality of a half-harmonic oscillator, the Hamiltonian~\eqref{initial_ham} can even be treated in the context of quantum optics, and might be explored in an experimental setting. We are hoping that the presented approach opens new avenues in the Hilbert-Pólya conjecture, especially in the Berry-Keating program.

\begin{acknowledgments}

The author acknowledges Andreas Deuchert, Mikhail Lemeshko, Douglas Lundholm, and Richard Schmidt for valuable discussions.

\end{acknowledgments}

\bibliography{bib_zeta.bib}


%

\end{document}